\documentclass[
reprint,
superscriptaddress,
amsmath,amssymb,
aps,prl]{revtex4-1}

\usepackage{graphicx}
\usepackage{dcolumn}
\usepackage{bm}
\usepackage{hyperref}
\usepackage[range-phrase={\,\text{--}\,}, range-units={single}, separate-uncertainty=true]{siunitx}

\newcommand{\ie}{i.\,e.\@}

\newcommand{\Ecm}{\ensuremath{E_\mathrm{c.\,m.}}}
\newcommand{\pg}{\ensuremath{({p},\gamma)}}
\newcommand{\pag}{\ensuremath{({p},\alpha\gamma)}}

\DeclareSIUnit\year{yr}
\DeclareSIUnit\solarMass{\ensuremath{\mathit{M}_\odot}}

\begin{document}


\title{
	\texorpdfstring
	{New proton-capture rates on carbon isotopes and their impact on the astrophysical $^{12}$C/$^{13}$C ratio}
	{New proton-capture rates on carbon isotopes and their impact on the astrophysical 12C/13C ratio}
}



\author{J.~Skowronski}
 \affiliation{Dipartimento di Fisica, Università degli Studi di Padova, 35131 Padova, Italy}
 \affiliation{INFN, Sezione di Padova, 35131 Padova, Italy}

\author{A.~Boeltzig}
 \email{a.boeltzig@hzdr.de}
 \affiliation{Helmholtz-Zentrum Dresden-Rossendorf, 01328 Dresden, Germany}
 \affiliation{INFN, Laboratori Nazionali del Gran Sasso, 67100 Assergi, Italy}
 \affiliation{Dipartimento di Fisica ``E. Pancini'', Universit\`a degli Studi di Napoli ``Federico II'', 80125 Naples, Italy}
 
\author{G.\,F.~Ciani}
 \affiliation{Dipartimento di Fisica ``M. Merlin'', Università degli Studi di Bari ``A. Moro'', 70125 Bari, Italy}
 \affiliation{INFN, Sezione di Bari, 70125 Bari, Italy}

\author{L.~Csedreki}
 \affiliation{Institute for Nuclear Research (ATOMKI), PO Box 51, H-4001 Debrecen, Hungary}

\author{D.~Piatti}
 \affiliation{Dipartimento di Fisica, Università degli Studi di Padova, 35131 Padova, Italy}
 \affiliation{INFN, Sezione di Padova, 35131 Padova, Italy}

\author{M.~Aliotta}
 \affiliation{SUPA, School of Physics and Astronomy, University of Edinburgh, EH9 3FD Edinburgh, United Kingdom}

\author{C.~Ananna}
 \affiliation{Dipartimento di Fisica ``E. Pancini'', Universit\`a degli Studi di Napoli ``Federico II'', 80125 Naples, Italy}
 \affiliation{INFN, Sezione di Napoli, 80125 Naples, Italy}

\author{F.~Barile}
 \affiliation{Dipartimento di Fisica ``M. Merlin'', Università degli Studi di Bari ``A. Moro'', 70125 Bari, Italy}
 \affiliation{INFN, Sezione di Bari, 70125 Bari, Italy}

\author{D.~Bemmerer}
 \affiliation{Helmholtz-Zentrum Dresden-Rossendorf, 01328 Dresden, Germany}

\author{A.~Best}
 \affiliation{Dipartimento di Fisica ``E. Pancini'', Universit\`a degli Studi di Napoli ``Federico II'', 80125 Naples, Italy}
 \affiliation{INFN, Sezione di Napoli, 80125 Naples, Italy}

\author{C.~Broggini}
 \affiliation{INFN, Sezione di Padova, 35131 Padova, Italy}

\author{C.\,G.~Bruno}
 \affiliation{SUPA, School of Physics and Astronomy, University of Edinburgh, EH9 3FD Edinburgh, United Kingdom}

\author{A.~Caciolli}
 \affiliation{Dipartimento di Fisica, Università degli Studi di Padova, 35131 Padova, Italy}
 \affiliation{INFN, Sezione di Padova, 35131 Padova, Italy}

\author{M.~Campostrini}
 \affiliation{Laboratori Nazionali di Legnaro, 35020 Legnaro, Italy}

\author{F.~Cavanna}
 \affiliation{INFN, Sezione di Torino, 10125 Torino, Italy}

\author{P.~Colombetti}
 \affiliation{Dipartimento di Fisica, Universit\`a degli Studi di Torino, 10125 Torino, Italy}
 \affiliation{INFN, Sezione di Torino, 10125 Torino, Italy}

\author{A.~Compagnucci}
 \affiliation{Gran Sasso Science Institute, 67100 L'Aquila, Italy}
 \affiliation{INFN, Laboratori Nazionali del Gran Sasso, 67100 Assergi, Italy}

\author{P.~Corvisiero}
 \affiliation{Università degli Studi di Genova, 16146 Genova, Italy}
 \affiliation{INFN, Sezione di Genova, 16146 Genova, Italy}

\author{T.~Davinson}
 \affiliation{SUPA, School of Physics and Astronomy, University of Edinburgh, EH9 3FD Edinburgh, United Kingdom}

\author{R.~Depalo}
 \affiliation{Università degli Studi di Milano, 20133 Milano, Italy}
 \affiliation{INFN, Sezione di Milano, 20133 Milano, Italy}

\author{A.~Di Leva}
 \affiliation{Dipartimento di Fisica ``E. Pancini'', Universit\`a degli Studi di Napoli ``Federico II'', 80125 Naples, Italy}
 \affiliation{INFN, Sezione di Napoli, 80125 Naples, Italy}

\author{Z.~Elekes}
 \affiliation{Institute for Nuclear Research (ATOMKI), PO Box 51, H-4001 Debrecen, Hungary}

\author{F.~Ferraro}
 \affiliation{INFN, Laboratori Nazionali del Gran Sasso, 67100 Assergi, Italy}
 \affiliation{Università degli Studi di Milano, 20133 Milano, Italy}

\author{A.~Formicola}
 \affiliation{INFN, Sezione di Roma, 00185 Roma, Italy}

\author{Zs.~Fülöp}
 \affiliation{Institute for Nuclear Research (ATOMKI), PO Box 51, H-4001 Debrecen, Hungary}

\author{G.~Gervino}
 \affiliation{Dipartimento di Fisica, Universit\`a degli Studi di Torino, 10125 Torino, Italy}
 \affiliation{INFN, Sezione di Torino, 10125 Torino, Italy}

\author{R.\,M.~Gesu\`e}
 \affiliation{Gran Sasso Science Institute, 67100 L'Aquila, Italy}
 \affiliation{INFN, Laboratori Nazionali del Gran Sasso, 67100 Assergi, Italy}

\author{A.~Guglielmetti}
 \affiliation{Università degli Studi di Milano, 20133 Milano, Italy}
 \affiliation{INFN, Sezione di Milano, 20133 Milano, Italy}

\author{C.~Gustavino}
 \affiliation{INFN, Sezione di Roma, 00185 Roma, Italy}

\author{Gy.~Gyürky}
 \affiliation{Institute for Nuclear Research (ATOMKI), PO Box 51, H-4001 Debrecen, Hungary}

\author{G.~Imbriani}
 \affiliation{Dipartimento di Fisica ``E. Pancini'', Universit\`a degli Studi di Napoli ``Federico II'', 80125 Naples, Italy}
 \affiliation{INFN, Sezione di Napoli, 80125 Naples, Italy}

\author{M.~Junker}
 \affiliation{INFN, Laboratori Nazionali del Gran Sasso, 67100 Assergi, Italy}

\author{M.~Lugaro}
 \affiliation{Konkoly Observatory, Research Centre for Astronomy and Earth Sciences (CSFK), MTA Centre for Excellence, 1121 Budapest, Hungary}
 \affiliation{ELTE E\"otv\"os Lor\'and University, Institute of Physics, 1117 Budapest, Hungary}

\author{P.~Marigo}
 \affiliation{Dipartimento di Fisica, Università degli Studi di Padova, 35131 Padova, Italy}
 \affiliation{INFN, Sezione di Padova, 35131 Padova, Italy}

\author{E.~Masha}
 \affiliation{Helmholtz-Zentrum Dresden-Rossendorf, 01328 Dresden, Germany}
 \affiliation{Università degli Studi di Milano, 20133 Milano, Italy}

\author{R.~Menegazzo}
 \affiliation{INFN, Sezione di Padova, 35131 Padova, Italy}

\author{V.~Paticchio}
 \affiliation{INFN, Sezione di Bari, 70125 Bari, Italy}

\author{R.~Perrino}
 \altaffiliation[Permanent address: ]{INFN Sezione di Lecce, 73100 Lecce, Italy}
 \affiliation{INFN, Sezione di Bari, 70125 Bari, Italy}

\author{P.~Prati}
 \affiliation{Università degli Studi di Genova, 16146 Genova, Italy}
 \affiliation{INFN, Sezione di Genova, 16146 Genova, Italy}

\author{D.~Rapagnani}
 \affiliation{Dipartimento di Fisica ``E. Pancini'', Universit\`a degli Studi di Napoli ``Federico II'', 80125 Naples, Italy}
 \affiliation{INFN, Sezione di Napoli, 80125 Naples, Italy}

\author{V.~Rigato}
 \affiliation{Laboratori Nazionali di Legnaro, 35020 Legnaro, Italy}

\author{L.~Schiavulli}
 \affiliation{Dipartimento di Fisica ``M. Merlin'', Università degli Studi di Bari ``A. Moro'', 70125 Bari, Italy}
 \affiliation{INFN, Sezione di Bari, 70125 Bari, Italy}

\author{R.\,S.~Sidhu}
 \affiliation{SUPA, School of Physics and Astronomy, University of Edinburgh, EH9 3FD Edinburgh, United Kingdom}

\author{O.~Straniero}
 \affiliation{INAF-Osservatorio Astronomico d'Abruzzo, 64100, Teramo, Italy}
 \affiliation{INFN, Sezione di Roma, 00185 Roma, Italy}

\author{T.~Szücs}
 \affiliation{Institute for Nuclear Research (ATOMKI), PO Box 51, H-4001 Debrecen, Hungary}

\author{S.~Zavatarelli}
 \affiliation{INFN, Sezione di Genova, 16146 Genova, Italy}
 \affiliation{Università degli Studi di Genova, 16146 Genova, Italy}

\collaboration{LUNA Collaboration}
\noaffiliation


\date{\today}


\begin{abstract}
The $^{12}$C/$^{13}$C ratio is a significant indicator of nucleosynthesis and mixing processes during hydrogen burning in stars. Its value mainly depends on the relative rates of the $^{12}$C$\pg{}^{13}$N and $^{13}$C\pg{}$^{14}$N reactions.
Both reactions have been studied at the Laboratory for Underground Nuclear Astrophysics (LUNA) in Italy down to the lowest energies to date ($\Ecm = \SI{60}{\kilo\electronvolt}$) reaching for the first time the high energy tail of hydrogen burning in the shell of giant stars.
Our cross sections, obtained with both prompt $\gamma$-ray detection and activation measurements, are the most precise to date with overall systematic uncertainties of \SIrange{7}{8}{\percent}.
Compared with most of the literature, our results are systematically lower, by \SI{25}{\percent} for the $^{12}$C$\pg{}^{13}$N reaction and by \SI{30}{\percent} for $^{13}$C$\pg{}^{14}$N. 
We provide the most precise value up to now of \num{3.6(4)} in the \SIrange{20}{140}{\mega\kelvin} range for the lowest possible $^{12}$C/$^{13}$C ratio that can be produced during H burning in giant stars.
\end{abstract}

\maketitle


Absorption lines in stellar spectra are commonly used to infer the atmospheric composition of a star. 
However, lines emitted by different isotopes of the same element can only be resolved for a limited number of cases. 
Among them, the carbon isotopic ratio $^{12}$C$/^{13}$C ratio provides important insights about internal nucleosynthesis and its coupling with various mixing process, such as those induced by thermal convection, rotational instabilities, thermohaline circulation, magnetic buoyancy and turbulence due to the propagation of gravity waves
\cite{dearborn1992, boothroyd1994, eleid1994, charbonnel1995, nollet2003, denissenkov2003, eggleton2008, palmerini2011, straniero2017}. Furthermore, the $^{12}$C$/^{13}$C ratio can be directly and precisely measured in silicon carbide (SiC) grains that originated in giant stars and are recovered from meteorites~\cite{zinner2014}.
In the solar system, $^{12}\mathrm{C}/{}^{13}\mathrm{C} = 89$~\cite{lodders2009}, while in nearby molecular clouds an average  $^{12}\mathrm{C}/{}^{13}\mathrm{C} \simeq 68$ has been reported~\cite{Langer1993, milam2005}. 
This lower ratio likely reflects the chemical evolution of the interstellar medium that occurred since the formation of the solar system \SI{4.6}{\giga\year} ago~\cite{milam2005}. The possible $^{13}$C producers that contribute to lowering the isotopic ratio include massive stars, Asymptotic Giant Branch (AGB) stars and classical novae.  
Measurements of  $^{12}\mathrm{C}/{}^{13}\mathrm{C}$ in the interstellar medium located at different galactocentric distances~\cite{milam2005} confirm that this ratio is a tracer of the chemical evolution: it is lower toward the galactic center, where the stellar density is higher and the chemical evolution faster, and higher away from the center where the stellar density is lower.

The carbon isotopic ratio in the atmospheres of evolved stars shows variations during their evolution that are the consequence of the combined action of internal nucleosynthesis and deep mixing processes. 
When a star, after the central H exhaustion, leaves the main sequence and becomes a red giant (RGB), a convective instability arises that starts from the surface and penetrates inwards, down into the region whose composition has been previously modified by H burning. 
As a result, $^{13}$C (but also $^4$He and $^{14}$N) is enhanced, while $^{12}$C is depleted and the atmospheric ratio is predicted to drop down to \SI{25(5)}{} (depending on the stellar initial composition and mass)~\cite{Boothroyd1999,Lebzelter2015}. 
This theoretical expectation is in good agreement with the abundances derived from optical and near infrared spectra, except for the brightest red giants with relatively low mass ($M < \SI{2.5}{\solarMass}$) that show a substantially lower isotopic ratio, \ie{}, $6 < {}^{12}\mathrm{C}/{}^{13}\mathrm{C} < 12$ \cite{gilroy1991, gratton2000, szigeti2018}. 
This may be due to the operation of an additional, non-convective, mixing process whose nature is currently under debate~\cite{Busso2007}. In principle there are several processes that may produce an extra-mixing below the inner boundary of the convective envelope of an RGB stars, among which, gravity waves, magnetic buoyancy and thermohaline circulation. 
Potential evidence for the occurrence of extra-mixing processes also exists for Asymptotic Giant Branch (AGB) and main-sequence massive stars~\cite{Aerts2014}. 
However, a fruitful exploitation of the carbon isotopic ratio as a probe of extra mixing requires a reliable evaluation of the stellar rates of both $^{12}$C\pg{}$^{13}$N \mbox{($Q=\SI{1.944}{\mega\electronvolt}$)} and  $^{13}$C\pg{}$^{14}$N \mbox{($Q=\SI{7.551}{\mega\electronvolt}$)} reactions.

In addition, the ${}^{12}$C\pg{}${}^{13}$N reaction  is one of the main sources of the solar CNO neutrino flux, recently observed in the Borexino experiment~\cite{borexino2020}, through the $\beta^{+}$ decay of $^{13}$N. 
The $^{12}$C($p,\gamma$) rate controls the onset of the CNO cycle, before it reaches equilibrium, and hence the radial profile of $^{13}$N neutrino emission~\cite{bahcall2006}.
The reported uncertainty of the newest reaction rate, \ie{}, NACRE2 compilation~\cite{xu2013}, is of the order of \SI{30}{\percent} in the temperature region of interest below \SI{0.02}{\giga K}, whereas a value of \SI{5}{\percent} is needed to constrain the solar neutrino fluxes~\cite{orebi2021}. We note that the older NACRE~\cite{angulo1999} compilation, widely used for astrophysical models, reports an uncertainty of the order of \SI{10}{\percent}, despite being based on the same experimental data as NACRE2.

At astrophysical energies, below \SI{120}{\kilo\electronvolt} in center of mass frame, the $^{12}$C\pg{}$^{13}$N and $^{13}$C\pg{}$^{14}$N are influenced by the low-energy tail of broad resonances, at $\Ecm \simeq \SI{421}{\kilo\electronvolt}$ and \SI{517}{\kilo\electronvolt} in the $^{12}$C\pg{}$^{13}$N and $^{13}$C\pg{}$^{14}$N reactions, respectively.  
The $^{12}$C\pg{}$^{13}$N reaction cross section was measured by both prompt $\gamma$-ray detection (\cite{vogl1963,rolfs1974,lamb1957,burtebaev2008}) and by the activation technique (\cite{baily1950,hall1950,artemov2016}), \ie{}, by detecting the decay radiation following the $\beta^{+}$ decay of $^{13}$N ($t_{\rm 1/2} = \SI{9.965(4)}{\minute}$~\cite{selove1991}).
An important discrepancy exists in the measured resonance energy between the two comprehensive studies by \citet{vogl1963} and \citet{rolfs1974}, the most recent of which dates back to 1974. 
In addition, at the lowest energies the large statistical uncertainties ($\geq \SI{10}{\percent}$) and data points scattering by \SI{30}{\percent}~\cite{baily1950,lamb1957} hinder a precise evaluation of the reaction rate.

The $^{13}$C\pg{}$^{14}$N reaction proceeds through radiative capture to five excited states in $^{14}$N and to its ground state. While several experiments were performed to measure the direct capture transition to the ground state~\cite{hester1961,woodbury1952,seagrave1952,vogl1963,genard2010}, only the study by \citet{king1994} reports the cross section for all transitions.
Although their cross section data extend down to \SI{100}{\kilo\electronvolt}, data points scatter by \SI{20}{\percent}, 
and show statistical uncertainties of the order of \SI{10}{\percent}. 
Thus a precise evaluation of the reaction rate at astrophysical energies remains challenging. 

Aiming to reconcile these discrepancies, we have performed an in-depth study of both reactions using the prompt $\gamma$-ray detection and, for the $^{12}$C\pg{}$^{13}$N reaction, also the activation technique.

The experiment was performed at the Laboratory for Underground Nuclear Astrophysics (LUNA) located in the Gran Sasso National Laboratories. 
The underground environment offers a remarkable sensitivity thanks to the background reduction afforded by \SI{1400}{\meter} of rock shielding~\cite{szucs2010}. 
The \SI{400}{\kilo\volt} LUNA accelerator provided a proton beam (up to \SI{400}{\micro\ampere}) in the energy range  $\Ecm = \SIrange{60}{370}{\kilo\electronvolt}$.
Three different types of targets were used: \SI{4}{\milli\meter} thick $^{\textup{nat}}$C disks, and $^{\textup{nat}}$C and \SI{99}{\percent} enriched $^{13}$C powders evaporated onto chemically cleaned Ta backings, which were produced and characterized at ATOMKI~\cite{ciani2019}. 
Their thickness was found to be \SIrange{8}{15}{\kilo\electronvolt} at the proton beam energy of \SI{380}{\kilo\electronvolt}. 
The thick $^{\textup{nat}}$C target was mainly used for the activation technique, but also served as a cross-check of data obtained with the evaporated targets. 
During beam bombardment, the target was cooled with deionized water. 
The entire target chamber was isolated and acted as a Faraday cup. 
A Cu pipe, extending to a few mm from the target, was cooled with LN$_2$ to act as a cold trap and was maintained at a potential of \SI{-300}{\volt} to suppress secondary electrons produced when the beam hits the target.

For both reaction studies, two complementary detection systems were used. 
The first involved the use of a \SI{120}{\percent} HPGe detector at an angle of \SI{0}{\degree} to the beam axis and at a distance of $\approx$ \SI{1.4}{\centi\meter} from the target.
The second detection system involved the use of a $4\pi$ BGO detector segmented into six crystals of equal shape and size~\cite{skowronski2023}. 
Given the much higher efficiency of the BGO compared with the HPGe, the second setup allowed us to push measurements to the lowest accessible energies ($\Ecm = \SI{60}{\kilo\electronvolt}$).
A \SI{15}{\centi\meter} (\SI{10}{\centi\meter}) thick lead castle was built around the target holder and the HPGe (BGO) detector in order to further reduce the environmental background at $E_{\gamma} < \SI{3}{\mega\electronvolt}$~\cite{boeltzig2018}. 
Two different DAQ systems were used: an analog chain with standard electronic modules to process the signal from the HPGe detector, and a CAEN V1724 digitizer to acquire energies and timestamps from the BGO crystals and to allow for coincidence summing of events from its segments. 
A pulser signal was also routed through the electronics of each DAQ to measure the dead time. 

The efficiency calibration curve of HPGe was obtained with a multiparametric minimization procedure~\cite{dileva2014} using data obtained with standard calibration sources and well-known resonances in the ${}^{14}$N\pg{}${}^{15}$O~\cite{imbriani2005} and ${}^{27}$Al\pg{}${}^{28}$Si~\cite{iliadis1990} reactions, so as to extend the efficiency curve up to $E_\gamma = \SI{8}{\mega\electronvolt}$.
Calibration data were acquired at four different target-detector distances to properly account for possible summing effects~\cite{gilmore2008} in the prompt-$\gamma$ measurements. 
The photo-peak detection efficiency at \SI{2}{\mega\electronvolt} and \SI{8}{\mega\electronvolt} was found to be \SI{2.76(6)}{\percent} and \SI{0.83(2)}{\percent}, respectively.

For the BGO setup, detection efficiencies were obtained from Geant4 simulations validated with  calibration sources and resonance measurements in the $^{27}$Al$\pg{}^{28}$Si and ${}^{14}$N\pg{}${}^{15}$O reactions. 
For the ${}^{13}$C\pg{}${}^{14}$N, we simulated all the different $\gamma$-ray cascades by using the observed decay branching ratios and obtained an efficiency of \SI{37(1)}{\percent} for the coincident detection of all the transitions that led to the ground state. 
For the ${}^{12}$C\pg{}${}^{13}$N we simulated the radioactive decay of $^{13}$N, leading to an efficiency of \SI{22.2(8)}{\percent} for the \SI{511}{\kilo\electronvolt} coincidences. 
The latter was validated with the use of the $^{14}$N$\pg{}^{15}$O reaction at \SI{279}{\kilo\electronvolt} resonance~\cite{imbriani2005}, where the produced $^{15}$O nuclei are $\beta^{+}$ unstable ($t_{1/2} = \SI{2.037(2)}{\minute}$~\cite{selove1991}). 
Additionally, the relative impact on the efficiency uncertainty of both beam spot position on target ($\simeq \SIrange{1}{2}{\percent}$) and  ${}^{13}$C\pg{}${}^{14}$N decay branching ratios  ($\simeq \SI{0.6}{\percent}$) were assessed with Monte Carlo techniques. 
 
Measurements with the HPGe detector were performed in the energy range $\Ecm = \SIrange{60}{370}{\kilo\electronvolt}$ in \SI{10}{\kilo\electronvolt} steps. 
Cross sections were extracted from a fit of the observed primary $\gamma$-ray peak shapes using the procedure described in \cite{dileva2014, ciani2020}. 
Briefly, because of the finite target thickness (\SIrange{25}{10}{\kilo\electronvolt} of beam energy loss for the entire energy range studied), the shape of the primary $\gamma$-ray peaks of the reaction of interest reflects a convolution of the beam energy loss due to the target thickness and stoichiometry (in the following referred to both as target profile) and the associated energy-dependence of the reaction cross section.
For the latter, we used resonance parameters available in the literature 
\cite{rolfs1974,king1994} as well as a non-resonant contribution included as a free parameter.

To monitor the target degradation yield measurements were repeated every \SI{10}{C} of accumulated charge on  target~\cite{ciani2020} at a reference energy $\Ecm \approx \SI{352}{\kilo\electronvolt}$. 
This energy was chosen to optimize  counting statistics while minimizing beam-induced background, mostly from a broad resonance at $\Ecm = \SI{322}{\kilo\electronvolt}$ in ${}^{19}$F\pag{}${}^{16}$O \cite{croft_absolute_1991}.
For most runs, we estimated an overall statistical uncertainty of \SI{1}{\percent} on the target profile and an overall systematic uncertainty of \SI{2.5}{\percent}.
Our results were compared with those obtained from target analyzes at ATOMKI~\cite{ciani2020}, performed after the irradiation, and found to be in agreement within \SI{2}{\percent}. 
We further checked for angular distribution effects by analyzing spectra taken with the HPGe detector at \SI{0}{\degree} and \SI{55}{\degree} and at an approximate distance of \SI{16.4}{\centi\meter} from the target to minimize summing effects. 
Our observations agreed very well with the expected isotropic distribution for $^{12}$C$\pg{}^{13}$N. 
In the case of $^{13}$C$\pg{}^{14}$N, a good agreement was found with results from \citet{king1994}, except for the very weak transition to the \SI{5105}{\kilo\electronvolt} state, for which our poor statistics prevented a definitive conclusion. 
Further, we compared yields obtained from HPGe measurements on evaporated targets with additional runs on $^{\textup{nat}}$C targets, finding an overall agreement within \SI{2}{\percent}.

For the BGO measurements, two different approaches were followed: prompt-$\gamma$-ray for the ${}^{13}$C\pg{}${}^{14}$N reaction, and mostly activation for the $^{12}$C$\pg{}^{13}$N reaction.
For ${}^{13}$C\pg{}${}^{14}$N reaction, 
measurements were performed at $\Ecm = \SIrange{60}{370}{\kilo\electronvolt}$  in \SI{10}{\kilo\electronvolt} steps;
$\gamma$-rays emitted in cascade from various transitions were seen in coincidence in all BGO crystals thanks to the nearly $4\pi$ coverage of the detector. 
By summing together signals from all individual transitions, we obtained a full absorption peak at $E_{\gamma} = Q + \Ecm \simeq \SI{7.8}{\mega\electronvolt}$, \ie{}, in a region unaffected by either the intrinsic background of the BGO crystals or the environmental radioactivity~\cite{boeltzig2018}. 
During BGO measurements, the target profile was regularly monitored by using the same $\gamma$-shape approach described above, for this purpose the BGO detector was retracted and the HPGe was positioned at \SI{55}{\degree} to the beam axis.

For the $^{12}$C$\pg{}^{13}$N reaction, measurements were performed on thick $^{\rm nat}$C targets at $\Ecm = \SIrange{70}{370}{\kilo\electronvolt}$ in steps of \SI{20}{\kilo\electronvolt} and an extra run at \SI{70}{\kilo\electronvolt}.
In this case, however, the prompt-$\gamma$ analysis was only possible at 
$\Ecm > \SI{185}{\kilo\electronvolt}$ due to the intrinsic BGO background. 
Instead, the activation measurement was done at all beam energies.
The \SI{511}{\kilo\electronvolt} annihilation $\gamma$-rays were detected in coincidence in opposite BGO crystals with negligible contribution from natural background events. 
The analysis was performed by fitting the \SI{511}{\kilo\electronvolt} coincidence rate, $\mathrm{d}N/\mathrm{d}t$, corrected by the detection efficiency, to the solution of the following differential equation:
\begin{align}
    \eta ^{-1} \frac{\mathrm{d}N}{\mathrm{d}t} = Y R_{p}(t) - \lambda N(t), \label{eq:activ:de}
\end{align}
where $\eta$ is the detection efficiency, $Y$ is the reaction yield (free parameter), $R_{p}(t)$ is the incoming proton rate, $\lambda$ is the $^{13}$N decay constant, and $N(t)$ is the number of $^{13}$N nuclei in the sample. 
For this purpose, the proton deposited charged was acquired in event per event basis, where one event equalled to 1 $\mu$C of accumulated charge. \cite{skowronski2023}.
The extracted yields were found to be in good agreement with those obtained from the prompt $\gamma$-ray approach in the common energy range.
Thick-target yields were finally differentiated from consecutive beam energy runs (\ie{}, \SI{20}{\kilo\electronvolt} steps) to obtain a thin-target yield, and thus extract the relevant cross section.

For all the measurements described above, cross section values were associated to effective interaction energies obtained as a cross section weighted average of the beam energy over the target thickness~\cite{iliadis2007}.
Cross section data were corrected for the electron screening effect following the procedure described in~\cite{assenbaum1987}. Corrections amounted up to \SI{12}{\percent} for the lowest energies.
We assumed a conservative uncertainty of $\SI{50}{\percent}$ on this correction. 

\begin{figure}[t]
    \centering
    \includegraphics[width=\linewidth]{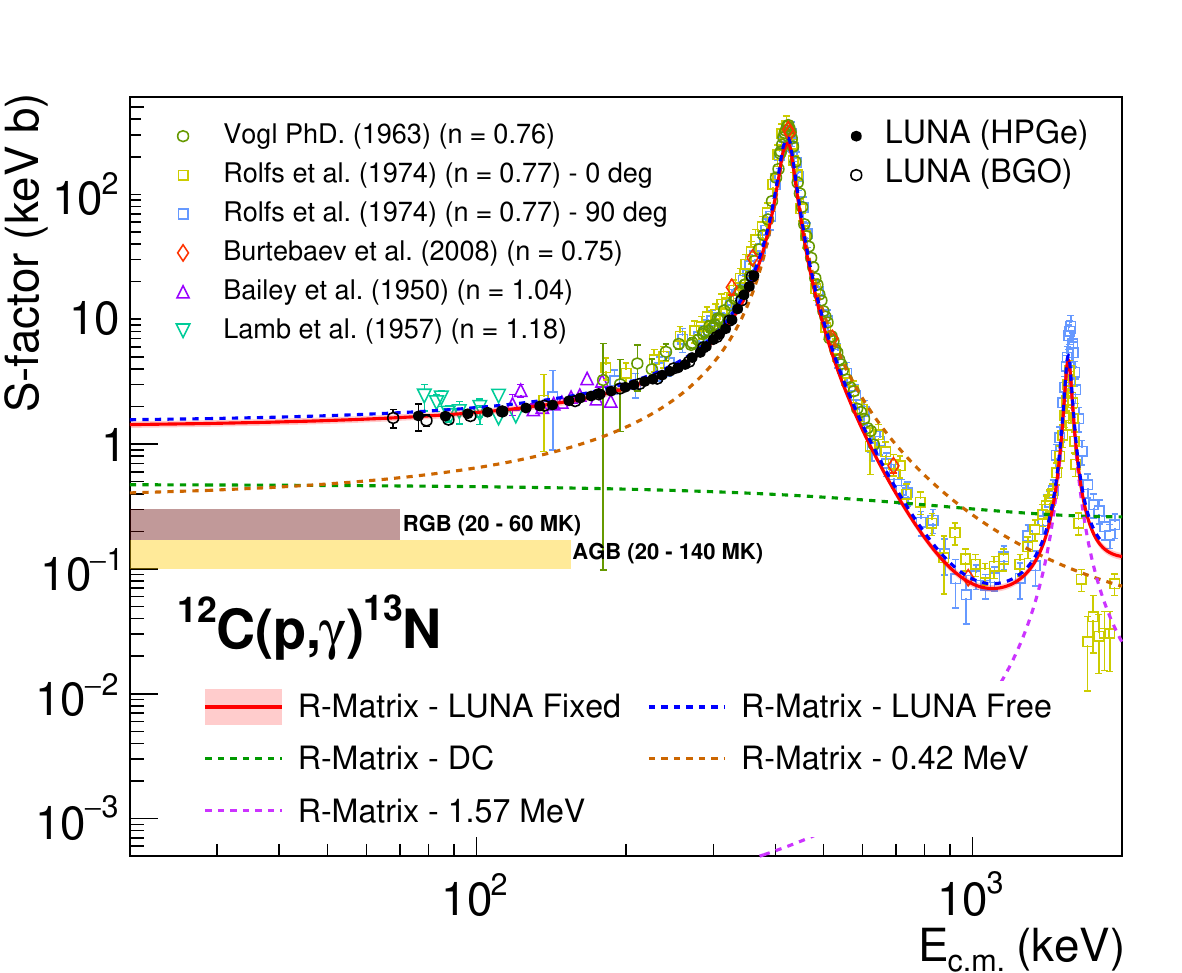}
    \includegraphics[width=\linewidth]{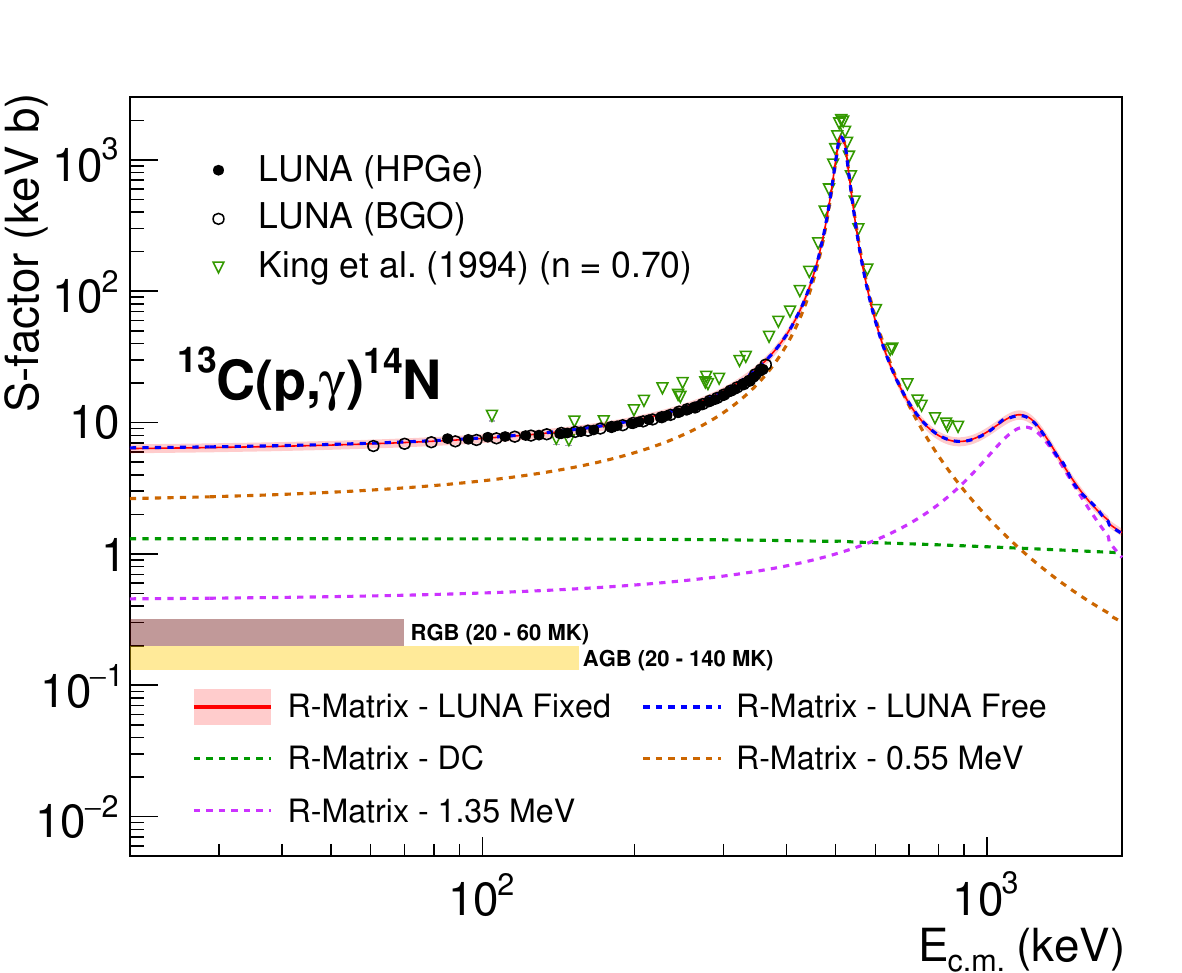}
    \caption{$R$-matrix fit to available $S$-factor data for the ${}^{12}$C\pg{}${}^{13}$N reaction (top panel) and the ${}^{13}$C\pg{}${}^{14}$N (bottom panel). The red solid line shows the best fit obtained with fixed LUNA data (black symbols) and normalized literature data (coefficients $n$ in the legend). The blue dashed line shows the fit obtained with free normalization also for the LUNA data. Uncertainties are statistical only. The brown and yellow regions represent the energy range relevant to RGB~\cite{salaris2002} and AGB stars~\cite{ventura2011,slemer2017}, respectively.
    }
    \label{fig:results:12c}
\end{figure}

Finally, our cross section data were converted into the astrophysical $S$-factor, $ S(E) = E \sigma(E) \exp[2\pi\eta(E)]$,
where $E$ is the energy in the center of mass frame and $\eta(E)$ is the Sommerfeld parameter~\cite{iliadis2007}. 
$S$-factor values for the two reactions are shown in Fig.~\ref{fig:results:12c} 
together with literature data, with error bars representing statistical uncertainties only ($\simeq \SI{1}{\percent}$ for our data, except for the lowest ones). Data sets obtained with the two different detection techniques are in excellent agreement with each other.
We note that our data extend down into  the hydrogen shell-burning energies (Gamow window) of AGB stars and, for the first time, of RGB stars.
Our more precise data are also important to better constrain the reaction rate extrapolations down to the typical temperature of core hydrogen burning in main sequence stars. 
By taking into account all the systematic uncertainties already mentioned, as well as a $\SI{6.4}{\percent}$ uncertainty associated with energy loss calculations based on SRIM~\cite{srim}, we estimated the following total systematic uncertainties:
$\SI{6.9}{\percent}$ (HPGe, prompt $\gamma$) and $\SI{7.9}{\percent}$ (BGO, activation) for the ${}^{12}$C\pg{}${}^{13}$N reaction; $\SI{7.2}{\percent}$ (HPGe, prompt $\gamma$) and $\SI{8.0}{\percent}$ (BGO, prompt $\gamma$) for the ${}^{13}$C\pg{}${}^{14}$N reaction.

For the ${}^{12}$C\pg{}${}^{13}$N reaction, our $S$-factors are consistently lower (by $\SI{25}{\percent}$) compared to most of the data in literature~\cite{vogl1963,rolfs1974,burtebaev2008}. 
For the ${}^{13}$C\pg{}${}^{14}$N reaction, a larger discrepancy (of about \SI{30}{\percent}) is observed compared to the results of \citet{king1994}.
The origin of such discrepancies is unclear, but may arise from an inaccurate treatment of target thickness effects in the literature, potentially affecting the extraction of cross sections from measured yields.
Indeed we note that the ${}^{12}$C\pg{}${}^{13}$N data by 
\citet{artemov2016}, presented in terms of reaction yields, are in agreement with our yields within \SI{4}{\percent}.

The extrapolation of $S$-factor data to the lowest energies of astrophysical interest was performed using the $R$-matrix formalism implemented in AZURE2~\cite{azuma2010}.
To evaluate the impact of the observed discrepancies on the extrapolation, we adopted two procedures:
in the first case, we fixed our LUNA $S$-factors and used normalization coefficients as free parameters for all other data sets; in
the second case, a free normalization coefficient was introduced also for our data set.
A systematic uncertainty of \SI{20}{\percent} was considered in the fit for all literature data sets that do not report any, apart from~\citet{king1994} and \citet{burtebaev2008}, who declare values of \SI{11.3}{\percent} and \SI{10}{\percent}, respectively.
The initial parameters for the $R$-matrix fit were taken from~\cite{artemov2022} and~\cite{chakraborty2015} for the proton capture on $^{12}$C and $^{13}$C, respectively.

The results of the fits are shown in Fig.~\ref{fig:results:12c}.
Red curves represent the best-fit with fixed LUNA data obtained with values of the normalization coefficients, $n$, as given in the labels.
The blue dashed lines correspond to best fits obtained with normalization coefficients of $1.07$ and $1.06$ for the present LUNA data, for the $^{12}$C\pg{}$^{13}$N and $^{13}$C\pg{}$^{14}$N reactions, respectively.
The difference between the two curves are well within the systematic uncertainty of the LUNA data sets. 
However, it also points out the need for further measurements of both reactions at energies around the resonance, with a more careful assessment of systematic uncertainties.

\begin{figure}[htbp]
    \centering
    \includegraphics[width=\linewidth]{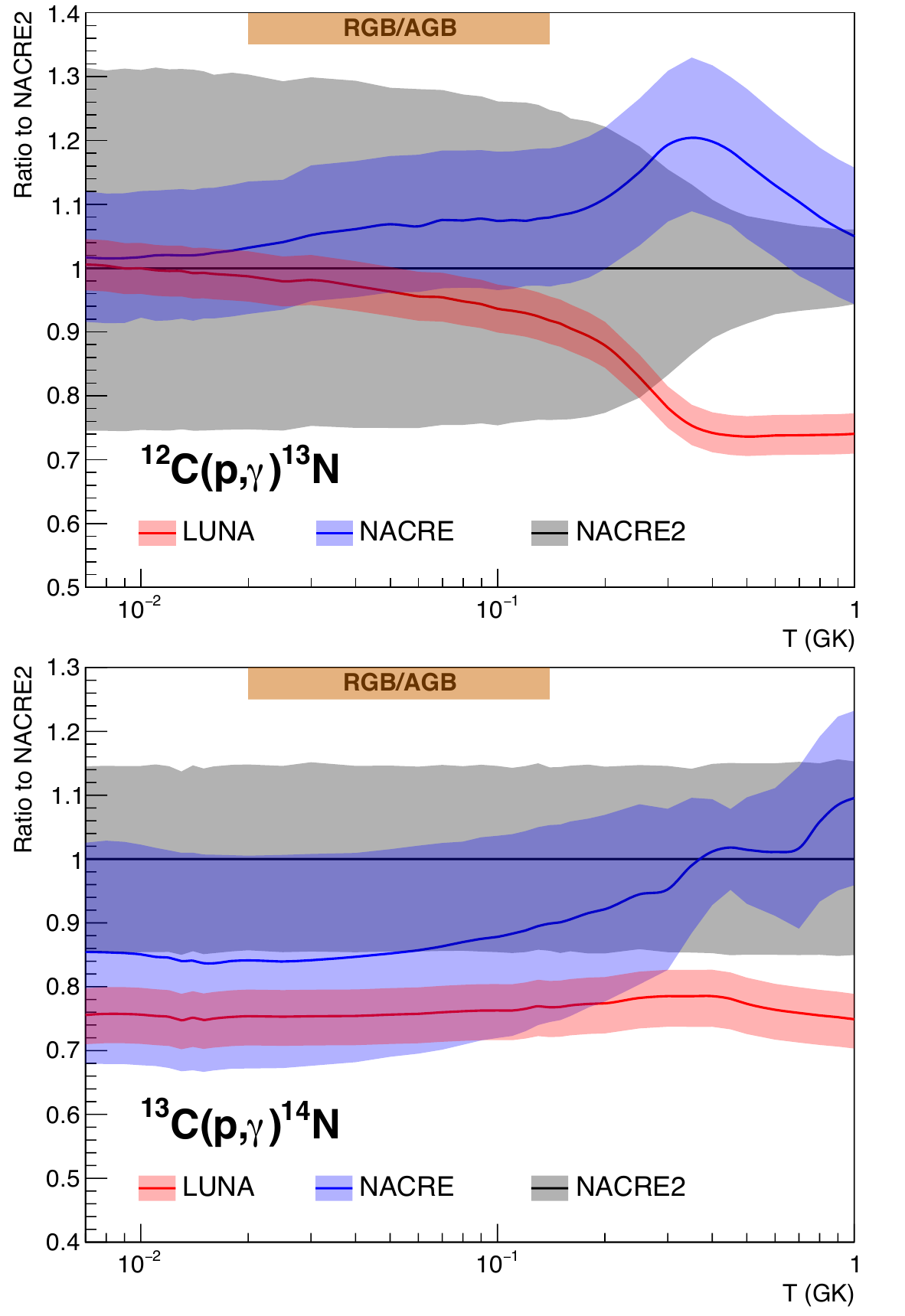}
    \caption{Reaction rate ratio to NACRE and NACRE2 values for the $^{12}$C$\pg{}^{13}$N (upper panel) and the ${}^{13}$C\pg{}${}^{14}$N (lower panel). The brown bar represents the temperature range of interest for RGB and AGB stars.}
    \label{fig:results:rates}
\end{figure}

The astrophysical rates for both reactions were calculated using the $R$-matrix extrapolations (red curves in Fig.~\ref{fig:results:12c}), as:
\begin{equation}
    \left< \sigma v \right>=\left ( \frac{8}{\pi\mu} \right )^{1/2}\left ( \frac{1}{kT} \right )^{3/2}\int_{0}^{\infty}E\sigma(E)e^{-{E/kT}}\mathrm{d}E
\end{equation}
where $\sigma$ and $\mu$ are the cross section and reduced mass for each reaction, $T$ is the stellar temperature, and $k$ the Boltzmann constant.
Our reaction rates are shown in Fig.~\ref{fig:results:rates} relative to the NACRE~\cite{angulo1999} and NACRE2~\cite{xu2013} compilations.
In the energy region of interest for RGB and AGB stars, our rates are significantly lower than both compilations for the ${}^{13}$C\pg{}${}^{14}$N reaction and are comparable in the ${}^{12}$C\pg{}${}^{13}$N case. The overall uncertainty ranges from \SIrange[]{4}{6}{\percent} in agreement with the one requested by the solar models~\cite{orebi2021}.

As mentioned, the observed abundance ratio of stable C isotopes is a significant tracer of galactic chemical evolution and a probe for the occurrence of non-canonical mixing processes in RGB, AGB and massive stars.
In H-burning layers, the $^{13}$C abundance is given by:
\begin{equation} \frac{\mathrm{d}n_{13}}{\mathrm{d}t}=n_{12}n_{p} \left< \sigma v \right>_{12} - n_{13}n_{p} \left< \sigma v\right>_{13}
\end{equation}
where $n_{p}$, $n_{12}$ and $n_{13}$ are the density of protons, $^{12}$C and $^{13}$C, respectively, and $\left< \sigma v \right>_{12}$ and $\left< \sigma v \right>_{13}$ are the production and destruction rates of $^{13}$C nuclei.
As usual, it is assumed that the $\beta^{+}$ decay of the $^{13}$N is much faster than the ${}^{12}$C\pg{}${}^{13}$N reaction. 
In the H-burning shell of a giant star, the $^{13}$C abundance shortly attains  equilibrium between production and destruction, \ie{}, $\mathrm{d}n_{13}/\mathrm{d}t = 0$.
This equilibrium condition implies that the C isotopic ratio is given by:
\begin{equation}
    \frac{n_{12}}{n_{13}}=\frac{\left< \sigma v \right>_{13}}{\left< \sigma v \right>_{12}}
\label{eq_4}
\end{equation}
which only depends on the temperature. 
In the \SIrange{20}{140}{\mega\kelvin} range, we obtain an average $n_{12}/n_{13} = \SI{3.6(4)}{}$, \ie{}, a lower and more precise value than $\SI{4(1)}{}$ and $\SI{5.0(13)}{}$ obtained with the NACRE compilations~\cite{angulo1999} and~\cite{xu2013}.
Our lower C-isotopic ratio alleviates the discrepancy between predicted and observed atmospheric composition in evolved RGB and AGB stars. 
In particular, our result requires a less extreme extra-mixing process to solve this discrepancy.  
A more detailed analysis of the astrophysical implications of our results based on detailed stellar models will be presented in a dedicated paper.

In summary, we measured the cross sections of the ${}^{12}$C\pg{}${}^{13}$N and ${}^{13}$C\pg{}${}^{14}$N reactions with the highest precision to date down to $\Ecm = \qty{68}{\kilo\electronvolt}$ and \qty{61}{\kilo\electronvolt}, respectively, \ie{}, within the Gamow energy region of interest for AGB and, for the first time, RGB stars.
Results obtained with different experimental techniques show an excellent agreement with each other. 
Statistical uncertainties are of the order of \SI{1}{\percent} for most of the data points (\SI{10}{\percent} below \SI{90}{\kilo\electronvolt}); systematic uncertainties are of the order of \SIrange{7}{8}{\percent}. 
Our $S$-factors are systematically lower than most of those in the literature and dominate $R$-matrix extrapolations at the lowest energies. 
Our revised reaction rates result in a reduced C isotopic ratio at relevant temperature for mixing effects in giant stars. 
More detailed evaluations, based on advanced stellar modeling, and the interpretation of observational constraints from stellar spectra and meteoritic stardust will be presented in forthcoming papers.

\smallskip

\begin{acknowledgments}
D. Ciccotti and the technical staff of the LNGS are gratefully acknowledged for their indispensable help.
Financial support by INFN,
the Italian Ministry of Education, University and Research (MIUR) through the ``Dipartimenti di eccellenza'' project ``Physics of the Universe'',
the European Union (ERC-CoG \emph{STARKEY}, no. 615604; ERC-StG \emph{SHADES}, no. 852016; and \emph{ChETEC-INFRA}, no. 101008324),
Deut\-sche For\-schungs\-ge\-mein\-schaft (DFG, BE~4100-4/1),
the Helm\-holtz Association (ERC-RA-0016),
the Hungarian National Research, Development and Innovation Office (NKFIH K134197, PD129060),
the European Collaboration for Science and Technology (COST Action ChETEC, CA16117)
is gratefully acknowledged.
M.A., C.G.B, T.D., and R.S.S. acknowledge funding from STFC (grant ST/P004008/1).
T.S. acknowledges support from the Bolyai research fellowship and by the New National Excellence Program of the Ministry of Human Capacities of Hungary (\'UNKP-22-5-DE-428).
J.S., G.I. and A.Bo. thank R.\,J.\ deBoer for his support with AZURE2.

\end{acknowledgments}

\providecommand{\noopsort}[1]{}\providecommand{\singleletter}[1]{#1}%

\end{document}


\title{\texorpdfstring{$^{12}$C($p,\gamma$)$^{13}$N and $^{13}$C($p,\gamma$)$^{14}$N Reaction Rates - Supplemental Material}
{12C(p,g)13N and 13C(p,g)14N Reaction Rates - Supplemental Material}}

\maketitle

The reaction rates for both, $^{12}$C(p,$\gamma$)$^{13}$N and $^{13}$C(p,$\gamma$)$^{14}$N, were calculated from the R-matrix extrapolations presented in the article.

\begin{table*}[ht]
\centering
\begin{minipage}{0.45\linewidth}
\caption{Thermonuclear reaction rate $N_{\rm A}\langle \sigma v\rangle$ for $^{12}$C(p,$\gamma$)$^{13}$N in units of cm$^{-3}$s$^{-1}$mol$^{-1}$, as a function of temperature $T_9$ in GK.}
\label{tab:rr:12c}
\begin{tabular}{| c | l | l | l |}
\hline
$T_9$ & Low Rate & Median Rate & High Rate \\
\hline
\hline
0.007 & $4.58 \times 10^{-23}$ & $4.77 \times 10^{-23}$ & $4.96 \times 10^{-23}$ \\
0.008 & $9.48 \times 10^{-22}$ & $9.88 \times 10^{-22}$ & $1.02 \times 10^{-21}$ \\
0.009 & $1.22 \times 10^{-20}$ & $1.27 \times 10^{-20}$ & $1.33 \times 10^{-20}$ \\
0.010  & $1.11 \times 10^{-19}$ & $1.15 \times 10^{-19}$ & $1.20 \times 10^{-19}$ \\
0.011 & $7.64 \times 10^{-19}$ & $7.96 \times 10^{-19}$ & $8.27 \times 10^{-19}$ \\
0.012 & $4.20 \times 10^{-18}$ & $4.38 \times 10^{-18}$ & $4.55 \times 10^{-18}$ \\
0.013 & $1.93 \times 10^{-17}$ & $2.01 \times 10^{-17}$ & $2.09 \times 10^{-17}$ \\
0.014 & $7.62 \times 10^{-17}$ & $7.94 \times 10^{-17}$ & $8.26 \times 10^{-17}$ \\
0.015 & $2.65 \times 10^{-16}$ & $2.76 \times 10^{-16}$ & $2.87 \times 10^{-16}$ \\
0.016 & $8.32 \times 10^{-16}$ & $8.66 \times 10^{-16}$ & $9.01 \times 10^{-16}$ \\
0.018 & $6.26 \times 10^{-15}$ & $6.52 \times 10^{-15}$ & $6.78 \times 10^{-15}$ \\
0.020  & $3.56 \times 10^{-14}$ & $3.71 \times 10^{-14}$ & $3.85 \times 10^{-14}$ \\
0.025 & $1.15 \times 10^{-12}$ & $1.20 \times 10^{-12}$ & $1.25 \times 10^{-12}$ \\
0.030  & $1.63 \times 10^{-11}$ & $1.70 \times 10^{-11}$ & $1.77 \times 10^{-11}$ \\
0.040  & $7.79 \times 10^{-10}$ & $8.11 \times 10^{-10}$ & $8.43 \times 10^{-10}$ \\
0.050  & $1.21 \times 10^{-8}$  & $1.26 \times 10^{-8}$  & $1.31 \times 10^{-8}$ \\
0.060  & $9.81 \times 10^{-8}$  & $1.02 \times 10^{-7}$  & $1.06 \times 10^{-7}$  \\
0.070  & $5.22 \times 10^{-7}$  & $5.43 \times 10^{-7}$  & $5.65 \times 10^{-7}$  \\
0.080  & $2.07 \times 10^{-6}$  & $2.16 \times 10^{-6}$  & $2.24 \times 10^{-6}$  \\
0.090  & $6.66 \times 10^{-6}$  & $6.94 \times 10^{-6}$  & $7.22 \times 10^{-6}$  \\
0.100   & $1.82 \times 10^{-5}$  & $1.90 \times 10^{-5}$  & $1.97 \times 10^{-5}$  \\
0.110  & $4.40 \times 10^{-5}$  & $4.59 \times 10^{-5}$  & $4.77 \times 10^{-5}$  \\
0.120  & $9.63 \times 10^{-5}$  & $1.00 \times 10^{-4}$  & $1.04 \times 10^{-4}$  \\
0.130  & $1.94 \times 10^{-4}$  & $2.02 \times 10^{-4}$  & $2.10 \times 10^{-4}$  \\
0.140  & $3.66 \times 10^{-4}$  & $3.81 \times 10^{-4}$  & $3.97 \times 10^{-4}$  \\
0.150  & $6.58 \times 10^{-4}$  & $6.81 \times 10^{-4}$  & $7.09 \times 10^{-4}$  \\
0.160  & $1.11 \times 10^{-3}$  & $1.15 \times 10^{-3}$  & $1.20 \times 10^{-3}$  \\
0.180  & $2.87 \times 10^{-3}$  & $2.99 \times 10^{-3}$  & $3.11 \times 10^{-3}$  \\
0.200   & $6.58 \times 10^{-3}$  & $6.86 \times 10^{-3}$  & $7.15 \times 10^{-3}$  \\
0.250  & $3.80 \times 10^{-2}$  & $3.96 \times 10^{-2}$  & $4.13 \times 10^{-2}$  \\
0.300   & $1.74 \times 10^{-1}$  & $1.81 \times 10^{-1}$  & $1.89 \times 10^{-1}$  \\
0.350  & $6.90 \times 10^{-1}$  & $7.19 \times 10^{-1}$  & $7.50 \times 10^{-1}$  \\
0.400   & $2.26 \times 10^{0}$   & $2.35 \times 10^{0}$   & $2.46 \times 10^{0}$   \\
0.450  & $6.09 \times 10^{0}$   & $6.35 \times 10^{0}$   & $6.63 \times 10^{0}$   \\
0.500   & $1.38 \times 10^{1}$   & $1.44 \times 10^{1}$   & $1.50 \times 10^{1}$   \\
0.600   & $4.80 \times 10^{1}$   & $5.01 \times 10^{1}$   & $5.22 \times 10^{1}$   \\
0.700   & $1.16 \times 10^{2}$   & $1.21 \times 10^{2}$   & $1.26 \times 10^{2}$   \\
0.800   & $2.21 \times 10^{2}$   & $2.31 \times 10^{2}$   & $2.41 \times 10^{2}$   \\
0.900   & $3.61 \times 10^{2}$   & $3.76 \times 10^{2}$   & $3.93 \times 10^{2}$   \\
1.000   & $5.27 \times 10^{2}$   & $5.50 \times 10^{2}$   & $5.73 \times 10^{2}$   \\
\hline
\end{tabular}
\end{minipage}
\hfill
\begin{minipage}{0.45\linewidth}
\caption{Thermonuclear reaction rate $N_{\rm A}\langle \sigma v\rangle$ for $^{13}$C(p,$\gamma$)$^{14}$N in units of  cm$^{-3}$s$^{-1}$mol$^{-1}$, as a function of temperature $T_9$ in GK.}
\label{tab:rr:13c}
\begin{tabular}{| c | l | l | l |}
\hline
$T_9$ & Low Rate & Median Rate & High Rate \\
\hline
\hline
0.007 & $1.66 \times 10^{-22}$ & $1.76 \times 10^{-22}$ & $1.86 \times 10^{-22}$ \\
0.008 & $3.46 \times 10^{-21}$ & $3.68 \times 10^{-21}$ & $3.88 \times 10^{-21}$ \\
0.009 & $4.49 \times 10^{-20}$ & $4.78 \times 10^{-20}$ & $5.05 \times 10^{-20}$ \\
0.010  & $4.09 \times 10^{-19}$ & $4.35 \times 10^{-19}$ & $4.59 \times 10^{-19}$ \\
0.011 & $2.82 \times 10^{-18}$ & $3.00 \times 10^{-18}$ & $3.16 \times 10^{-18}$ \\
0.012 & $1.55 \times 10^{-17}$ & $1.65 \times 10^{-17}$ & $1.74 \times 10^{-17}$ \\
0.013 & $7.16 \times 10^{-17}$ & $7.62 \times 10^{-17}$ & $8.04 \times 10^{-17}$ \\
0.014 & $2.83 \times 10^{-16}$ & $3.02 \times 10^{-16}$ & $3.18 \times 10^{-16}$ \\
0.015 & $9.90 \times 10^{-16}$ & $1.05 \times 10^{-15}$ & $1.11 \times 10^{-15}$ \\
0.016 & $3.10 \times 10^{-15}$ & $3.30 \times 10^{-15}$ & $3.49 \times 10^{-15}$ \\
0.018 & $2.34 \times 10^{-14}$ & $2.49 \times 10^{-14}$ & $2.63 \times 10^{-14}$ \\
0.020  & $1.33 \times 10^{-13}$ & $1.42 \times 10^{-13}$ & $1.50 \times 10^{-13}$ \\
0.025 & $4.36 \times 10^{-12}$ & $4.64 \times 10^{-12}$ & $4.90 \times 10^{-12}$ \\
0.030  & $6.20 \times 10^{-11}$ & $6.61 \times 10^{-11}$ & $6.97 \times 10^{-11}$ \\
0.040  & $2.96 \times 10^{-9}$  & $3.15 \times 10^{-9}$  & $3.32 \times 10^{-9}$ \\
0.050  & $4.61 \times 10^{-8}$  & $4.90 \times 10^{-8}$  & $5.17 \times 10^{-8}$ \\
0.060  & $3.73 \times 10^{-7}$  & $3.97 \times 10^{-7}$  & $4.19 \times 10^{-7}$  \\
0.070  & $1.98 \times 10^{-6}$  & $2.11 \times 10^{-6}$  & $2.22 \times 10^{-6}$  \\
0.080  & $7.86 \times 10^{-6}$  & $8.37 \times 10^{-6}$  & $8.83 \times 10^{-6}$  \\
0.090  & $2.52 \times 10^{-5}$  & $2.68 \times 10^{-5}$  & $2.83 \times 10^{-5}$  \\
0.100   & $6.87 \times 10^{-5}$  & $7.32 \times 10^{-5}$  & $7.71 \times 10^{-5}$  \\
0.110  & $1.65 \times 10^{-4}$  & $1.76 \times 10^{-4}$  & $1.85 \times 10^{-4}$  \\
0.120  & $3.60 \times 10^{-4}$  & $3.83 \times 10^{-4}$  & $4.04 \times 10^{-4}$  \\
0.130  & $7.22 \times 10^{-4}$  & $7.69 \times 10^{-4}$  & $8.10 \times 10^{-4}$  \\
0.140  & $1.35 \times 10^{-3}$  & $1.44 \times 10^{-3}$  & $1.52 \times 10^{-3}$  \\
0.150  & $2.40 \times 10^{-3}$  & $2.55 \times 10^{-3}$  & $2.69 \times 10^{-3}$  \\
0.160  & $4.06 \times 10^{-3}$  & $4.32 \times 10^{-3}$  & $4.55 \times 10^{-3}$  \\
0.180  & $1.03 \times 10^{-2}$  & $1.09 \times 10^{-2}$  & $1.15 \times 10^{-2}$  \\
0.200   & $2.31 \times 10^{-2}$  & $2.46 \times 10^{-2}$  & $2.59 \times 10^{-2}$  \\
0.250  & $1.19 \times 10^{-1}$  & $1.27 \times 10^{-1}$  & $1.34 \times 10^{-1}$  \\
0.300   & $4.49 \times 10^{-1}$  & $4.79 \times 10^{-1}$  & $5.04 \times 10^{-1}$  \\
0.350  & $1.45 \times 10^{0}$   & $1.55 \times 10^{0}$   & $1.63 \times 10^{0}$  \\
0.400   & $4.40 \times 10^{0}$   & $4.68 \times 10^{0}$   & $4.93 \times 10^{0}$   \\
0.450  & $1.21 \times 10^{1}$   & $1.29 \times 10^{1}$   & $1.36 \times 10^{1}$   \\
0.500   & $3.00 \times 10^{1}$   & $3.20 \times 10^{1}$   & $3.37 \times 10^{1}$   \\
0.600   & $1.29 \times 10^{2}$   & $1.37 \times 10^{2}$   & $1.44 \times 10^{2}$   \\
0.700   & $3.79 \times 10^{2}$   & $4.03 \times 10^{2}$   & $4.25 \times 10^{2}$   \\
0.800   & $8.50 \times 10^{2}$   & $9.06 \times 10^{2}$   & $9.54 \times 10^{2}$   \\
0.900   & $1.58 \times 10^{3}$   & $1.68 \times 10^{3}$   & $1.77 \times 10^{3}$   \\
1.000   & $2.57 \times 10^{3}$   & $2.74 \times 10^{3}$   & $2.88 \times 10^{3}$   \\
\hline
\end{tabular}
\end{minipage}
\end{table*}